\documentclass[preprint2]{aastex}
\received{June 5, 2000}
\accepted{August 30, 2000}

\slugcomment{To be published in the Astrophysical Journal, Vol. 546, Jan 10
2001}

\shorttitle{X-rays from Pictor A}
\shortauthors{Wilson, Young \& Shopbell}

\begin{document}

\title{Chandra X-ray Observations of Pictor A: High Energy Cosmic Rays
in a Radio Galaxy?}

\author{A. S. Wilson\altaffilmark{1}, A. J. Young and P. L.
Shopbell\altaffilmark{2}}

\affil{Astronomy Department, University of Maryland, College Park,
MD 20742; wilson@astro.umd.edu, ayoung@astro.umd.edu, pls@astro.umd.edu}

% Notice that each of these authors has alternate affiliations, which
% are identified by the \altaffilmark after each name.  The actual alternate
% affiliation information is typeset in footnotes at the bottom of the
% first page, and the text itself is specified in \altaffiltext commands.
% There is a separate \altaffiltext for each alternate affiliation
% indicated above.

\altaffiltext{1}{Adjunct Astronomer, Space Telescope Science Institute,
3700 San Martin Drive,
Baltimore, MD 21218; awilson@stsci.edu}

\altaffiltext{2}{Present address: Department of Astronomy, Mail Code 105-24,
California Institute of Technology, Pasadena, CA 91125}

% The abstract environment prints out the receipt and acceptance dates
% if they are relevant for the journal style.  For the aasms style, they
% will print out as horizontal rules for the editorial staff to type
% on, so long as the author does not include \received and \accepted
% commands.  This should not be done, since \received and \accepted dates
% are not known to the author.

\begin{abstract}
We report X-ray observations of the nearby, powerful radio galaxy Pictor A with
the Chandra Observatory and optical and near uv 
observations of its western radio 
hot spot with the Hubble Space Telescope. X-ray emission is detected from the
nucleus, a 1\farcm9 (110 kpc) long jet to the west of the nucleus, the
western radio hot spot some 4\farcm2 (240 kpc) from the nucleus, and the
eastern radio lobe. The morphology of the western hot spot is remarkably
similar to that seen at radio and optical wavelengths, where the emission is
known to be synchrotron radiation. The X-ray spectrum of the hot spot is well
described by an absorbed power law with photon index 
$\Gamma$ = 2.07 $\pm$ 0.11. 
The X-ray jet coincides with a weak radio jet and is laterally extended
by $\simeq$ 2\farcs0 (1.9 kpc). The observed jet is up to $\simeq$
15 times brighter
in X-rays than any counter jet, a difference ascribed to relativistic boosting 
as the western  radio lobe is probably the closer. The jet's spectrum is
well modelled by an absorbed power law with $\Gamma$ = 1.94$^{+0.43}_{-0.49}$ 
and
poorly fitted by a Raymond-Smith thermal plasma model.

The emission processes responsible for the X-rays are discussed in detail.
The radio-optical spectrum of the hot spot breaks or turns down at 10$^{13-14}$
Hz, and its X-ray spectrum is not a simple extension of the radio-optical
spectrum to higher frequencies.
Thermal models for the hot spot's X-ray emission are ruled
out.
Synchrotron self-Compton models involving
scattering from the {\it known} population of electrons give the
wrong spectral index for the hot spot's X-ray emission and are also excluded.
A composite synchrotron plus synchrotron self-Compton model can match
the X-ray observations but requires similar contributions from the two
components in the Chandra band. We show that the hot spot's X-ray emission 
could be synchrotron self-Compton emission from a hitherto unobserved
population of electrons emitting at low radio frequencies, but do not favor this
model in view of the very weak magnetic field required. 

An inverse Compton model of the jet, in which it scatters microwave background
photons but moves non-relativistically, requires a magnetic field a factor of
$\simeq$ 30 below
equipartition, and ad hoc conditions to explain why the radio lobes are
fainter than the jet in X-rays but brighter in the radio. These problems
are alleviated if the jet moves relativistically, but models with
an equipartition field require an implausibly
small angle ($\theta$) between the jet and the line of sight. 
A model with $\theta$ $\simeq$ 23$^{\circ}$ and a field
a factor of 6 below equipartition seems viable.

Synchrotron radiation is an alternative process for the X-ray emission. The
expected synchrotron spectrum from relativistic electrons accelerated by strong
shocks and subject to synchrotron radiation losses is in very good agreement
with that observed for both the hot spot and jet. The possibility that the
relativistic electrons result via photo-pion production by high energy
protons accelerated in shocks
(a `proton induced cascade') is briefly discussed.
\end{abstract}

% The different journals have different requirements for keywords.  The
% keywords.apj file, found on aas.org in the pubs/aastex-misc directory, 
% contains a list of keywords used with the ApJ and Letters.  These are 
% usually assigned by the editor, but authors may include them in their
% manuscripts if they wish. 


% That's it for the front matter.  On to the main body of the paper.
% We'll only put in tutorial remarks at the beginning of each section
% so you can see entire sections together.

% In the first two sections, you should notice the use of the LaTeX \cite
% command to identify citations.  The citations are tied to the
% reference list via symbolic KEYs.  We have chosen the first three
% characters of the first author's name plus the last two numeral of the
% year of publication.  The corresponding reference has a \bibitem
% command in the reference list below.
%
% Please see the AASTeX manual for a more complete discussion on how to make
% \cite-\bibitem work for you.   

\newpage
\section{Introduction}

X-ray observations of radio galaxies are of value for several reasons.
Low brightness radio lobes may be thermally confined by the hot,
extended atmospheres in elliptical galaxies or clusters of galaxies
and X-ray observations can provide the density, temperature and
pressure of this gas, thus constraining the pressures of the
relativistic gas and magnetic field in a lobe (e.g. Worrall \&
Birkinshaw 2000). High brightness regions, such as the hot spots in
Fanaroff-Riley class II (FRII) radio galaxies, cannot be statically
confined by thermal pressure and must be expanding and/or advancing
away from the nucleus.  If the motion is supersonic, such a feature
must be preceded by a bow shock in the surrounding medium. The effect
of bow shocks on the ambient gas may be identifiable in thermal X-ray
emission, depending on the run of temperature and density in the
post-shock gas (e.g. Clarke, Harris \& Carilli 1997). Thermal X-rays may also
be detectable from radio jets
and lobes themselves if they entrain and shock surrounding gas (e.g.
de Young 1986).  The relativistic particles responsible for the
synchrotron radio emission must also radiate through inverse Compton
scattering of ambient photons (such as the microwave background and
the starlight of the galaxy) or the synchrotron photons themselves.
This inverse Compton radiation is expected to be spread over a wide
range of frequencies, including the X-ray band.  Detection of both
synchrotron and inverse Compton radiation provides a direct
measurement of the magnetic field in the emitting region (e.g. Harris,
Carilli \& Perley 1994), allowing a check on the common assumption of
equipartition of energy between cosmic rays and magnetic fields.
Lastly, synchrotron X-ray emission may be observable if electrons
and/or positrons of sufficiently high energy are present, a finding
which would provide important clues about the origin of the
relativistic particles in these objects. For these reasons, we have
begun a program of imaging and spectroscopy of radio galaxies with the
Chandra X-ray Observatory.

Pictor A is the seventh brightest extragalactic radio source in the
sky at 408 MHz (Robertson 1973\footnote{Robertson's catalog excludes
  Cyg A on account of its low galactic latitude}), and the most
powerful radio galaxy (P$_{408 MHz}$ = 6.7 $\times$ 10$^{26}$ W
Hz$^{-1}$ for z = 0.035, q$_{0}$ = 0 and H$_{0}$ = 50 km s$^{-1}$
Mpc$^{-1}$) with z $<$ 0.04.  The best radio maps (Perley, R\"oser \&
Meisenheimer 1997, hereafter PRM) reveal two round, diffuse lobes with
very much brighter radio hot spots on the sides away from the nucleus,
corresponding to an FRII classification.  The total angular diameter
of the radio emission is about 7\farcm6 (430 kpc). The western hot
spot, some 4\farcm2 (240 kpc) from the nucleus,
is a remarkable object, being amongst the brightest of radio hot
spots in radio galaxies and quasars.  PRM have found a very faint
radio jet extending from the compact, flat spectrum (cf. Jones, McAdam
\& Reynolds 1994) nuclear core to the western hot spot.  Recently,
VLBI observations (Tingay et al. 2000) have revealed a milli arc
second- (pc-) scale nuclear radio jet which is approximately aligned
with the large-scale radio jet.  Simkin et al. (1999) have reported an
optical continuum extension $\simeq$ 0\farcs1 (95 pc) to the W of the
nucleus and the direction to this extension is also roughly aligned
with the larger scale radio jet.

R\"oser \& Meisenheimer (1987) discovered optical emission from the
western hot spot. This emission has a featureless continuum, is
strongly polarized and thus of synchrotron origin. These authors also
found a marginal detection of the hot spot at X-ray wavelengths in an
Einstein IPC observation. Further observations of the hot spot in the
infrared have been reported by Meisenheimer, Yates \& R\"oser (1997).
The overall spectrum of the western hot spot conforms to a power law
with index $\alpha$ = 0.74 (S $\propto$ $\nu^{-\alpha}$) from 327 MHz
to $\sim$ 10$^{14}$ Hz, where the spectrum turns down (see Fig. 2i of
Meisenheimer, Yates \& R\"oser 1997).  The morphologies of the hot
spot are very similar at radio and optical wavelengths, with a bright,
compact, leading ``core'' and a fainter, following ``filament'', which
extends $\simeq$ 15$^{\prime\prime}$ (14 kpc) more or less
perpendicular to a line joining the hot spot to the nucleus (PRM).
Fainter radio emission associated with
the hot spot is also seen to the E and NE of the optical and radio
filament, the overall angular extent of the radio hot spot being
$\simeq$ 25$^{\prime\prime}$ (24 kpc). The radio to optical spectrum
of the ``filament'' is somewhat steeper ($\alpha_{opt}^{3.6cm}$ =
0.98) than that of the bright, compact region ($\alpha_{opt}^{3.6cm}$
= 0.87). Optical imaging and polarimetry of the bright part of the hot
spot with the Hubble Space Telescope (Thomson, Crane \& Mackay 1995)
confirms the high polarization ($\gtrsim$ 50\%) and resolves the hot
spot into highly polarized ``wisps'' elongated nearly perpendicular to
the nucleus - hot spot line.

The nucleus of Pictor A has long (Schmidt 1965; Danziger, Fosbury \&
Penston 1977) been known for its strong emission-line spectrum, with
broad wings to both forbidden and permitted lines. Double-peaked, very
broad Balmer lines were discovered in 1993/1994 (Halpern \& Eracleous
1994; Sulentic et al. 1995); these double-peaked lines were not
present in 1983. Such line profiles are commonly interpreted as
emission from an accretion disk (e.g. Storchi-Bergmann et al. 1997).
The nucleus of Pictor A is a strong X-ray source; the spatially
integrated emission observed with ASCA has a power-law continuous
spectrum and no evidence for an Fe K$\alpha$ line (Eracleous \&
Halpern 1998). Further observations of the X-ray spectrum of Pictor A
have been reported recently by Padovani et al. (1999) and Eracleous,
Sambruna \& Mushotzky (2000).

We selected Pictor A for observation with Chandra because of the X-ray
emission from the western hot spot, tentatively detected with the
Einstein IPC and confirmed by us through an inspection of an archival
ROSAT PSPC observation. Further, the optical synchrotron emission from
the hot spot is of sufficient extent ($\sim$ 15$^{\prime\prime}$) to
be well resolved by Chandra, making this object a prime candidate for
the study of high energy processes in the hot spots of powerful radio
galaxies. In addition to the Chandra observations, we also report
analysis of archival Hubble Space Telescope (HST) optical and near uv
imaging observations of the western hot spot.

Section 2 describes the observations and their reduction while Section
3 presents the results. In Section 4, we discuss the orientation of the radio
source and the processes responsible for the X-ray emission of the western
hot spot and jet. Concluding remarks are given in Section 5.

\section{Observations and Reduction}

\subsection{Chandra X-ray Observations}

Pictor A was observed by the Chandra X-ray Observatory on Jan 18 2000
(sequence number 700018, obsid 346) using the Advanced CCD Imaging
Spectrometer (ACIS) spectroscopic array, which provides excellent
spatial resolution ($\le$ 1$^{\prime\prime}$) with medium
spectroscopic resolution ($\simeq$ 130 eV FWHM at 1 keV) at the aim
point on chip S3 (back illuminated). In order to obtain good spatial
resolution on the western hot spot while retaining excellent
resolution on the nucleus, the latter was placed about 1$^{\prime}$
from the aim point in the direction towards the S3/S2 chip boundary
(the --Y direction). The nucleus was thus about 1$^{\prime}$ from the
S3/S2 boundary. The western hot spot was $\simeq$ 3\farcm5 from the
aim point and also on chip S3. The spatial resolution at the nucleus is
barely different from the optimal, while that at the hot spot is
1\farcs2 $\times$ 1\farcs9 FWHM. The eastern radio lobe is partly on
S3 and partly on S2 (front illuminated). The total integration time
(``Good Time Intervals'') was 26.177 ksecs, taken with the default
frame time of 3.2 sec. There were no flares in the background count
rate, which remained constant at 1.4 counts per second over the S3
chip.

The data were processed using version R4CU4UPD4.2 of the pipeline
software, and we have used the updated gain file
acisD1999-09-16gainN0003.fits. The data extraction and analysis have
been performed using version 1.1 of the CIAO software and version 11.0
of XSPEC.

In order to create an X-ray image covering the whole radio source
(shown as Fig. 1), it was necessary to remove CCD artifacts that take
the form of broad, linear features along the readout direction in chips
S2 and S3. This was accomplished by first copying a spatial subsection
of the level 2 events file and rotating it by -37.6$^{\circ}$, so that the axis
of the S array ran horizontally. The image was then smoothed with a Gaussian
with $\sigma$ = 1\farcs0 and the average of the rows computed, including
rejection of those pixels in each column which were more than 2$\sigma$
from the mean for that column. The average of the rows was then subtracted
from each row. Finally, the image was rotated by +37.6$^{\circ}$, so the
rows and columns corresponded to the cardinal directions. This process
not only removed most of the linear artifacts, but also the `streak' from the
very strong nuclear source that results from its readout and much of the
differences in background level between the two chips.

To derive the X-ray spectrum of the hot spot, counts were extracted
from a rectangular region $15\arcsec \times 20\arcsec$ in extent which
includes all of the X-ray emission from the hot spot. Background
counts were taken from an area of identical shape offset to the west.
Other extraction regions were tried for the background but no
significant changes were seen. We obtain a total of 3413 counts from
the hot spot and 125 from the background. Response and auxiliary
response files were created using the standard CIAO tools.  The source
counts were grouped so that each channel has at least 25 counts so
that the $\chi^2$ fitting in XSPEC is meaningful.

We followed a similar procedure to obtain the X-ray spectrum of the
jet.  Counts were taken from a rectangular region 103$^{\prime\prime}$
$\times$ 2\farcs4 that contains the whole jet and minimises background
(the $\sim$ 10$^{\prime\prime}$ gap between the nucleus and the
eastern end of the jet was excluded).  The background was taken from
adjacent regions, and no significant difference was found for the
jet's spectrum when using different regions for the background. We
obtain a total of 481 counts from the jet and 240 from the background.
The X-ray response of the CCDs depends on the location of the source
on the chip. To account for this, the back-illuminated S3 chip has
1024 individual calibration measurements (256 per node) arranged in a
$32\times32$ grid covering the face of the chip. This oversamples the
variation in the response of the CCD, and a single calibration
measurement may be used for a source extending up to $64\times64$
pixels (approximately 32$^{\prime\prime}$ $\times$
32$^{\prime\prime}$).  The X-ray jet extends 228 pixels ($1\farcm9$)
from the nucleus, so we divided it into four sections, each of which
has a separate response. We also note that the jet crosses a `node
boundary' (the division between sections of the chip that have
different read-out amplifiers), and that the response may vary sharply
across this boundary. To avoid any problem with this, we also divided
the jet at the node boundary so as not to mix events collected by
different nodes.

Each of the four spectra were grouped to give each channel at least 20
counts so that $\chi^2$ fitting is valid. These spectra were fitted
simultaneously in XSPEC with each spectrum having its own background,
response and auxiliary response file. The spectral properties are
assumed to be constant along the length of the jet so the parameters
in the fit are tied together with the exception of their
normalizations.

One drawback of treating the jet in this way is that the individual
spectra will have extremely low count rates, and fitting may only be
valid over a limited range in energy. It is also difficult to produce
clear figures of the spectral fits since there are four almost
coincident data sets. To overcome these problems the entire jet
spectrum may be treated as a single data set, using an `averaged'
response function that is the photon weighted average of the four
individual responses. We use PI spectra, which are gain
corrected, to minimize the effect of collecting photons from two
different nodes. The results we obtain using this technique are
equivalent, within the errors, to those obtained through fitting the
spectra individually.

\subsection{HST Optical and Near UV Observations}

Six images of the western hot spot were obtained on August 19, 1995 as
part of GO program \#5931 (PI Meisenheimer), three each in the F300W
and F622W broadband filters.  The images in each filter were dithered
by several arc seconds in order to improve the effective spatial
resolution of the images on the PC chip.  However, the lack of
duplicate exposures at each pointing, i.e., no CR-SPLIT, and the
paucity of sources in the field greatly complicated attempts at
alignment of the images.  Instead, a number of cosmic rays were
cleaned from each image by hand, and then the total number of
sky-subtracted counts in the hot spot was determined for each frame.
After averaging the results for each filter, the counts were converted
to units of flux using the SYNPHOT package in IRAF\footnote{IRAF is
distributed by the National Optical Astronomy Observatories, which
are operated by the Association of Universities for Research in
Astronomy, Inc., under cooperative agreement with the National
Science Foundation.}.  The derived values, as plotted in Fig. 6, are
30~$\mu$Jy and 104~$\mu$Jy for the F300W and F622W filters,
respectively.  The errors in these values have been estimated at 10\%.

\section{Results} 

\subsection{Overall Morphology}

A low resolution X-ray image, covering all of the radio emitting
regions and with selected VLA radio contours (from PRM) overlaid,
is shown in Fig. 1, while a full resolution X-ray image of
the nucleus and structure to the west of the nucleus is shown in
Fig. 2. The brightest X-ray source is associated with the nucleus of
the galaxy. The position of this source, as measured with the Chandra
aspect system, is within 2$^{\prime\prime}$ of the radio and optical
nuclear positions given by PRM. The nucleus is so bright in X-rays
that it suffers from strong ``pile up'', so no useful spatial or
spectroscopic information can be obtained from this exposure; a
Chandra observation with a shorter frame time is scheduled to obtain a
reliable nuclear spectrum.
A linear feature, visible in both figures, extends
1\farcm9 (110 kpc) westward from the nucleus and is spatially
coincident with the faint radio jet (PRM); we shall refer to this
feature as the ``X-ray jet''. The X-ray jet ``points'' at the western
hot spot, which is 4\farcm2 (240 kpc) from the nucleus. When the
nuclear X-ray source is shifted slightly to coincide with the nuclear
radio source, the peak of the X-ray emission from the hot spot is
within 1$^{\prime\prime}$ of its radio peak. There is also evidence for faint,
diffuse X-ray emission around the nucleus and extending
$\sim$ 1$^{\prime}$ from it, but more sensitive observations are needed to
confirm this component.

Faint X-ray emission extends eastwards from the nucleus to
the two eastern radio hot spots, 3\farcm1 and 3\farcm4 from the
nucleus (Fig. 1). 
This faint X-ray emission is seen to brighten somewhat within
$\sim$ 1$^{\prime}$ of these hot spots. There are also four compact X-ray
sources within the brightest radio contour around the eastern hot spots;
none of these X-ray sources coincides with either radio hot spot.
Lastly, we point out a faint
($40-50$ counts above the background), compact X-ray source
323$^{\prime\prime}$ E of the nucleus at $\alpha$ = 05$^{h}$ 20$^{m}$
20$^{s}$.0, $\delta$ = --45$^{\circ}$ 47$^{\prime}$
44$^{\prime\prime}$ (J2000). This X-ray source is beyond the eastern
radio lobe, but it lies in p.a. 101$^{\circ}$ from the nucleus, a
direction which is precisely opposite to the direction to the western
hot spot (p.a. 281$^{\circ}$).  Examination of the Digital Sky Survey
reveals a faint point source within $\sim$ 1$^{\prime\prime}$
of this position, almost
at the limit of the SERC-J Survey (we estimate $m_{\rm
  SERC-J}\approx21$), and marginally detected in the Second Epoch
Southern (UK Schmidt) survey. It is unclear whether this compact X-ray
source is related to Pictor A.

\subsection{The Western Hot Spot}

\subsubsection{Morphology}

Images of the western hot spot at radio (PRM), optical (R\"oser 1989;
PRM) and X-ray wavelengths with similar resolutions are shown in
Fig. 3.  As described in the Introduction, the radio and optical hot
spots have a very similar morphology. In a higher resolution
(0\farcs45 $\times$ 0\farcs09) radio image (Fig. 19 of PRM), the bright
``core'' is found to have a sharp leading (i.e. away from the nucleus)
edge, with a slower decline on the side towards the nucleus. Behind
this bright region is a linear feature elongated in p.a. 32$^{\circ}$
(Fig. 3),
referred to as the ``filament'' by R\"oser \& Meisenheimer (1987) and the
``plateau'' by PRM.

The overall X-ray morphology is remarkably similar to the radio and optical,
with
both the ``core'' and the ``filament'' detected.
Much of the apparent elongation of the brighter part of the ``core'' of
the X-ray hot spot (Fig. 3) reflects the shape of the point spread function
(psf). Fig. 4 compares profiles along the major and minor axes of the hot spot
through both the hot spot and the model psf at the location of the hot spot.
At half maximum, the hot spot is somewhat wider than the 
model, but the model psf does not include broadening by imperfections in
the aspect solution. In order to assess the reliability of the model psf,
we have compared 
a compact
X-ray source (visible in Fig. 1) 
some 152$^{\prime\prime}$ from the nucleus in p.a.
40$^{\circ}$ with the Chandra model psf at this location. The
observed profile of this source (which is at a similar distance from
the aim point as the hot spot) is in good agreement with the model psf, but
broader by 0\farcs2 at the half power point. This number is similar to the
excess broadening of the hot spot over the model psf along its minor axis
(Fig. 4), suggesting that the brighter part of the hot spot is unresolved
in this direction. The excess FWHM of the hot spot over the psf along its
major axis (Fig. 4) is somewhat larger. 
Nevertheless, we feel that caution is necessary, since other effects (e.g. the 
motion of the source on the detector during an observation) could lead to
further broadening compared with the model psf. The conservative conclusion
is that the brightest part of the ``core'' is marginally resolved or
unresolved in X-rays.

There are, however, clear indications of structure in the hot spot at lower
brightness levels. As in the radio, the
X-ray core has a sharper edge away from the nucleus than towards it. A
narrow
``ridge'' 
extends $\simeq$ 3\farcs5 (3.4 kpc) in p.a. 108$^{\circ}$
(the general direction towards the nucleus) from the hot spot peak.
A weaker feature some 6$^{\prime\prime}$ (5.7 kpc) SSW of the core and
elongated towards p.a. 279$^{\circ}$, is seen in both the radio and
X-ray images. In conclusion, when the small differences in resolution
of the X-ray and radio images (Fig. 3) are taken into account,
the X-ray and radio morphologies are remarkably similar. This morphological
similarity indicates that the X-ray, optical and radio-emitting regions are
physically co-spatial and argues that the X-ray emission does not originate
from a bow shock ahead of the radio hot spot.

\subsubsection{X-ray Spectrum}

The X-ray spectrum of the hot spot 
has been modelled in various ways. A hot plasma
with solar abundances and in collisional equilibrium
(``Raymond-Smith'' model) provides a poor fit to the data (Table 1)
and can be ruled out. If the abundances of all elements heavier than
helium are allowed to vary, we find that an abundance of $\lesssim$ 10
per cent solar is needed for this thermal model to provide an
acceptable description of the data. However, this model (Table 1)
requires a hydrogen column well below the Galactic value of N$_{H}$ =
4.2 $\times$ 10$^{20}$ atoms cm$^{-2}$ towards Pictor A (Heiles \&
Cleary 1979), and is thus implausible. Fixing the column at the
Galactic value while allowing the abundances to vary leads to a poor
fit (Table 1). Further, we shall argue in Section 3.2.3 that a thermal
model can be excluded on grounds that the required gas density is so
high that the radio emission would be Faraday depolarized, contrary to
observation. Lastly, a power law of photon index $\Gamma$ =
2.07$^{+0.11}_{-0.11}$ absorbed by a column density of solar abundance
gas N$_{H}$ = (7.1$^{+2.3}_{-2.3}$) $\times$ 10$^{20}$ atoms cm$^{-2}$
provides an excellent description of the spectrum (Table 1, Fig. 5).
This column density is only slightly greater than the Galactic column
density towards Pictor A, suggesting that the interstellar medium of
our Galaxy is responsible for most of the observed absorption.
The unabsorbed flux and luminosity of the western hot spot
in the 2 - 10 keV band
are 3.1 $\times$ 10$^{-13}$ erg cm$^{-2}$ s$^{-1}$ and 1.7 $\times$ 10$^{42}$
erg s$^{-1}$, respectively.

\subsubsection{X-ray Emission Mechanism}

We first evaluate the thermal model. If we assume a representative
intrinsic dimension of 1.9 $\times$ 0.9 $\times$ 0.9 kpc (i.e.
2$^{\prime\prime}$ $\times$ 1$^{\prime\prime}$ $\times$
1$^{\prime\prime}$), as suggested for the bright ``core'' by the X-ray
image (Figs 3 and 4), the required gas density for the best fitting
Raymond-Smith model is n$_{e}$ $\simeq$ 3 cm$^{-3}$. 
Because the X-rays come from the same general region
of space as the radio and optical emission (Fig. 3), we may estimate
the magnetic field in this putative thermal gas from the synchrotron
emission. For an equipartition magnetic field of 4.7 $\times$
10$^{-4}$ gauss (obtained assuming a total cosmic ray energy equal to
twice that of the electrons, $\alpha$ = 0.74 [Meisenheimer, Yates \& R\"oser
1997], lower and upper cut-off
frequencies 10$^{7}$ and 10$^{14}$ Hz, respectively, and an emitting
volume corresponding to the size of the radio core - 760 $\times$ 190
$\times$ 190 pc [0\farcs8 $\times$ 0\farcs2 $\times$ 0\farcs2]), the
rotation measure through the hot spot is expected to be $\simeq$ 6
$\times$ 10$^{5}$ rad m$^{-2}$. If the larger volume adopted above for
the X-ray source is used to calculate the equipartition field, a value
of 1.4 $\times$ 10$^{-4}$ gauss and a rotation measure through the hot
spot of $\simeq$ 2 $\times$ 10$^{5}$ rad m$^{-2}$ are found.  In
contrast, the high polarization observed at 6 cm implies a rotation
measure internal to the hot spot $<$ 900 rad m$^{-2}$ (n$_{e}$ $<$
0.014 cm$^{-3}$), a factor of $\simeq$ 200 lower than required for
thermal emission. If the 20 cm polarization is used, the upper limit
on rotation measure and density are an order of magnitude smaller, but
the hot spot is not well resolved at this wavelength. We thus conclude
that our upper limit to the gas density renders a thermal model
untenable. This conclusion could be wrong if either a) the magnetic
field has a preferred direction but many reversals; such a field
structure can provide the high observed synchrotron polarization but
gives little Faraday rotation, or b) the thermal gas is actually in
small, dense clumps with a small covering factor. We favor neither of
these: the field structure of a) is physically implausible, and it is
very improbable that thermal gas of density 3 cm$^{-3}$, let alone the
higher density required if the gas is clumped, exists 240 kpc from the
Pictor A galaxy. Another argument for a low gas density in the hot
spot comes from the photoelectrically absorbing column. Most or all of
this column comes from gas in our Galaxy; any contribution from other
regions is $<$ 3 $\times$ 10$^{20}$ atoms cm$^{-2}$ (Table 1). If
spread uniformly throughout the hot spot ``core'' of average diameter
$\sim$ 1 kpc, the density is $<$ 0.1 cm$^{-3}$ (as long as electrons
are not stripped from the K shells of the relevant elements), a factor
of 30 below the density needed for thermal emission. In summary, our
finding (Section 3.2.2) that a thermal model provides a poor
description of the X-ray spectrum, and the upper limits to the gas
density from the limits on Faraday depolarization and the absorbing
column, rule out a thermal model of the X-rays from the hot spot
We conclude that the X-rays are non-thermal in origin and discuss
relevant models in Section 4.2.2.

\subsection{The X-ray Jet}

\subsubsection{Morphology}

The X-ray jet extends 1\farcm9 (110 kpc) westward from the nucleus in
p.a. 281$^{\circ}$ (Fig. 2).  This direction is only 1$^{\circ}$ away
from that quoted for the radio jet out to 3$^{\prime}$ (170 kpc) from
the nucleus (PRM). Such a difference in p.a. is within the errors of
measurement, so we can conclude that the X-ray and radio jets are
coincident.

The profile of X-ray emission along the jet, shown in Fig.~6, consists
of a number of `knots', the brightest of which is $30\arcsec$ from the
nucleus. In Section 4.1, we shall argue that the western radio lobe is
the nearer.
The ratio of the X-ray
fluxes of the western and eastern jets, using
the brightest part of the western jet, is more than a factor of
10 (Fig. 6), suggesting relativistic
boosting, as discussed in
Section 4.3.2.

We have compared the transverse profiles across the jet with the
expected psf. After deconvolution from the psf, the jet is found to be
transversely extended. The width of the jet varies somewhat along its
length, but its typical FWHM = 2\farcs0 (1.9 kpc). We are confident of
this transverse extent because: a) the profile of the compact source
152$^{\prime\prime}$ from the nucleus in p.a. 40$^{\circ}$ agrees well
with the expected psf at its location (Section 3.2.1), and b) the
linear feature that results from the CCD readout of the
strong nuclear source gives a measure of the 1d psf at the nucleus;
this feature is much narrower than the jet.  These arguments show that
the measured jet width is not a result of poor aspect solution or
focus.

The jet is so faint in the radio that it is difficult to measure its
width.  PRM state ``no accurate estimate of the jet's width is
possible, but it is clearly not greatly resolved by the 7\farcs5
beam''. Thus the upper limit to the radio width is consistent with the
X-ray width.

\subsubsection{X-ray spectrum}

Given the low count rate, the background begins to dominate above 2.5 keV.
The same models as used for the hot spot were tried for the jet (Table
2). 
The Raymond-Smith thermal models provide a poor fit to the
spectrum, even when the metal abundance is a free parameter.  An
absorbed power-law model provides an excellent description of the
spectrum of the jet (Fig. 7).  The absorbing column density ($N_{\rm
H}$ = 5.8$^{+6.8}_{-5.6}$ $\times$ 10$^{20}$ atoms cm$^{-2}$) and
photon index ($\Gamma$ = 1.94$^{+0.43}_{-0.49}$) are similar to those
found for the hot spot. Again we conclude that most of the absorption
occurs in the interstellar medium of our Galaxy. The
unabsorbed 2 - 10 keV flux
and luminosity of the jet are 2.5 $\times$ 10$^{-14}$ erg cm$^{-2}$ s$^{-1}$
and 1.4 $\times$ 10$^{41}$ erg s$^{-1}$, respectively.

\subsubsection{X-ray Emission Mechanism}

Even if the metal abundance is left as a free parameter,
a Raymond-Smith thermal plasma model provides a poor fit to the
jet's spectrum (Section 3.3.2, Table 2). 
Since
there is no information on the radio polarization of the jet, we
cannot argue against a thermal model on grounds of internal Faraday
depolarization, as we did for the western hot spot. The X-ray
absorbing column within the jet must be $\lesssim$ 8.4 $\times$ 10$^{20}$
atoms cm$^{-2}$ (the maximum allowed column [Table 2] minus the
Galactic column), which implies a density n$_{\rm{e}}$ $<$ 0.15
cm$^{-3}$. For the observed X-ray emission to be thermal, the average
density of hot gas in the jet (taken to be a cylinder of radius
1$^{\prime\prime}$ and length 1\farcm9) is n$_{\rm{e}}$ $\simeq$ 0.05 cm$^{-3}$,
compatible with the limits from the absorbing column. Thus the only argument
against thermal emission is the poor description of the spectrum by a 
Raymond-Smith model. We will discuss non-thermal models for the jet
in Section 4.3.2.

\section{Discussion} 
\subsection{Orientation of the Radio Source}

The projected distance, d$_{\rm{w}}$,
of the W radio hot spot
of Pictor A from the nucleus is larger than that, d$_{\rm{e}}$,
of the E radio
hot spots.
This observation is consistent with the W radio lobe being the nearer
one, since the extra light travel time to the more distant lobe implies that
it is seen at an earlier phase in its motion away from the nucleus and thus
appears closer to the nucleus.
Further, the east lobe is more depolarized than the west
at long radio wavelengths 
(PRM), again suggesting that the W lobe is the
nearer (Laing 1988).
Both the radio and X-ray jets of Pictor A are seen on the W side of the
nucleus. These considerations strongly suggest that the jet sidedness is a
result of relativistic boosting.

The angle of the source axis to the line of sight is, of course,
unknown, but can be estimated from d$_{\rm{w}}$ and d$_{\rm{e}}$ if the
source is assumed to be intrinsically symmetric, since

\begin{equation}
\frac{d_{\rm{w}} - d_{\rm{e}}}{d_{\rm{w}} + d_{\rm{e}}} = \beta\cos\theta
\end{equation}

\noindent
where $\beta$ = V/c, V is the velocity with which the hot spots are receding
from the nucleus and $\theta$ is the angle between the radio axis and our line
of sight. A recent study (Arshakian \& Longair 2000) has found the mean
value of $\beta$ in a sample of FRII radio galaxies to be 0.11 $\pm$ 0.013.
For Pictor A, we may then write

\begin{equation}
\cos \theta = 0.92\ \left (\rm{{V}\over{0.11 c}}\right)^{-1}
\end{equation}

\noindent
so $\theta$ = 23$^{\circ}$ if V = 0.11 c. In view of its sensitivity to V
and the assumption of symmetry, this value of $\theta$ should not be taken too
seriously.

The radio lobes are remarkably round and are well separated in the
VLA images (PRM), although joined by a `waist' near the nucleus.
The overall radio extent in the plane of the sky is 430 kpc and
the outward motions of the VLBI jet components (Tingay et al. 2000) are
subluminal.
All these properties indicate that $\theta$ is not small.
A large value of $\theta$
is suggested by the fact that the hot spots project further away from
the nucleus than the outer edges of the lobes themselves. Tingay et al. (2000)
have estimated that the VLBI-observed jet has $\theta$ $<$ 51$^{\circ}$ based
on a jet-deflection model. 

\subsection{The X-ray Emission of the Western Hot Spot}

\subsubsection{Radio to X-ray Spectrum}

Fig.~8 shows the broad band spectrum of the western hot spot. All of
the radio points and most of the optical and near infrared ones were
taken from Meisenheimer, Yates \& R\"oser (1997). To these we added
the flux densities at $\simeq$ 2900 \AA\ and $\simeq$ 6200 \AA\ 
obtained from our analysis of archival HST data (Section 2.2). These
HST values may underestimate the true flux densities as some of the
emission is resolved out.  Also included is the original estimate of
the X-ray flux density from the Einstein Observatory (R\"oser \&
Meisenheimer 1987) and the Chandra-measured intrinsic spectrum. The Einstein
point is a factor of $\sim$ 2 higher than the Chandra measurement; this may
be a result of partial blending of the hot spot with the much stronger
nuclear source in the low spatial resolution Einstein observation. The
radio spectrum of the hot spot is well described by a power law with
$\alpha$ = 0.740 $\pm$ 0.015 (Meisenheimer, Yates \& R\"oser 1997),
but there must be a break or turnover in the spectrum
at 10$^{13-14}$ Hz to accommodate the near infrared and optical
measurements. It is also apparent that the X-ray spectrum is not a
simple extension of the radio and optical measurements to higher
frequencies. The luminosities of the radio to optical and the X-ray
components are given in Table 3.

\subsubsection{Non-Thermal Models}
\subsubsubsection{Inverse Compton and Synchrotron Emission from the Radio and
Optical-Emitting Electron Population}

It has long been realised that the outward
motion of hot spots in FRII radio galaxies is sub-relativistic. The most recent
work (Arshakian \& Longair 2000) obtains a mean outward velocity
of 0.11c $\pm$ 0.013c. While the velocity of a hot spot in
any individual object is
uncertain, we shall neglect boosting or diminution
of the hot spot fluxes by bulk relativistic motion.

The broad-band spectrum (Fig. 8) shows that the X-ray emission of the
western hot spot cannot be synchrotron radiation from a high energy
extension of the population of electrons responsible for the radio to
optical synchrotron radiation. It is, therefore, natural to consider
inverse Compton scattering for the X-ray emission. In the bright core
of the hot spot, the radiant energy density ($\epsilon_{rad}$) of the
synchrotron radiation dominates that of the microwave background
radiation and the galaxy starlight (Table 3), so it is appropriate to
consider a synchrotron self-Compton model\footnote{In the
  ``filament'', the energy densities of the synchrotron radiation and
    the microwave background are comparable.}.  When the scattering
    electrons follow a power-law distribution in $\gamma$ (= E/mc$^{2}$),
n($\gamma$) = n$_{e_0}\gamma^{-\rm{p}}$ over all $\gamma$, then the
    inverse Compton spectrum is a power law with a spectral index
    $\alpha_C = (\rm{p}-1)/2$, the same index as the synchrotron
    spectrum.\footnote{Because we are concerned with only scattered
photons below an energy of 10 keV, we assume Thomson scattering
throughout i.e. $\gamma h \nu_{S} \ll m c^{2}$, where $\nu_{S}$ is
the frequency of the synchrotron photon and $m c^{2}$ is the rest
mass energy of the electron.}  The fact that the radio
($\alpha_{rad}$ = 0.740 $\pm$ 0.015) and X-ray ($\alpha_{X-ray}$ =
1.07$^{+0.11}_{-0.11}$) spectral indices are different then suggests
difficulties with an inverse Compton model.  However, in reality, the
electron energy spectrum is a power law over only a certain range of
$\gamma$ (from $\gamma_{min}$ to $\gamma_{max}$), and the synchrotron
spectrum is a power law over only a certain range of $\nu$ (from
$\nu_{s,min}$ to $\nu_{s,max}$). These limited ranges yield ``end
effects'' in synchrotron self-Compton spectra: the inverse Compton
scattered spectra must turn down below $\nu_{c,min} \sim
4\gamma^{2}_{min} \nu_{s,min}$ and above $\nu_{c,max} \sim
4\gamma^{2}_{max} \nu_{s,max}$ (e.g. Blumenthal \& Gould 1970; Rybicki
\& Lightman 1979). Further, the synchrotron self-Compton spectrum is
no longer an exact power law between these limits.

In order to provide a more realistic evaluation of inverse Compton and
synchrotron self-Compton models, we have performed numerical
calculations of spectra in spherical geometries using the computer
code of Band \& Grindlay (1985, 1986), which was kindly provided by
Dan Harris. Given the observed radio~--~optical spectrum (Fig. 8), we
have assumed a power law electron spectrum with $p$ = 2.48 from
$\gamma_{min}$ to $\gamma_{max}$. The magnetic field is treated as a
variable. We first computed a synchrotron self-Compton spectrum which
passes through the Chandra-measured flux (model 1). This was achieved
for a magnetic field strength of $3.3\times10^{-5}$ gauss, a factor of
14 below the equipartition field in the radio ``core'' of the hot spot
(Table 4).  However, the predicted spectrum is similar to that of the
radio source and does not match the measured X-ray spectrum (Fig.  8).

As an alternative to reducing the field strength, the power radiated in
inverse Compton radiation could be increased by increasing the radiation
density. The nucleus of Pictor A might emit a narrow, collimated beam of
radiation (like those inferred to be present in BL Lac objects) along the axis
of the jet. This beam could illuminate the hot spot, but be invisible to us.
However, an (isotropic) nuclear luminosity of 1.6 $\times$ 10$^{48}$
erg s$^{-1}$ is needed merely to {\it equal} the radiant energy density
in synchrotron radiation in the core of the hot spot. The (isotropic)
nuclear luminosity would have to be $\simeq$ 1.5 $\times$ 10$^{50}$ erg
s$^{-1}$ to provide sufficient radiation for the hot spot
to radiate
the observed X-ray flux by inverse Compton scattering if the field has
its equipartition value. This luminosity is unreasonably large.

An alternative is that the X-rays are a combination of synchrotron and inverse
Compton radiation.
In model 2, we assumed that the turnover in the synchrotron spectrum
at 10$^{13-14}$ Hz is an effect of synchrotron losses on a
continuously injected electron spectrum with energy index $p$ = 2.48.
The assumed electron spectrum is thus a broken power law (Kardashev 1962):
\begin{eqnarray}
n_e=n_{e_0}\gamma^{-p} & \gamma \ll\ \gamma_{\rm break} \\
n_e=n_{e_0}\gamma_{\rm break}\gamma^{-(p+1)} & 
\gamma \gg\ \gamma_{\rm break}
\end{eqnarray}
\noindent
The synchrotron spectrum then has an index $\alpha$ = 0.74 well below the
break and $\alpha$ = 0.74 + 0.5 = 1.24 is expected well above it
(dotted line in Fig.
8).  By adding this spectrum to the predicted synchrotron self-Compton
component for a magnetic field of $5.3\times10^{-5}$ gauss (dot-dashed
line in Fig. 8), which is again well below equipartition, we obtain
the solid line in Fig. 8, which is a good description of the Chandra
spectrum.  The parameters of model 2 are given in Table 4.
Because the energy density in magnetic field is $\gtrsim$ an order of
magnitude larger than the synchrotron 
radiation density, the power in the first order
synchrotron self-Compton component is less than that in the synchrotron
radiation (e. g. Rees 1967). Further, the second order scattered component
is even weaker (3 orders of magnitude below the first order radiation in the
Chandra band).

Assuming that the electrons are continuously accelerated in the hot
spot, we may interpret the turnover frequency as the frequency at which
the ``half-life'' to synchrotron losses of the radiating electrons is equal
to their
escape time from the hot spot. For the field of $5.3\times10^{-5}$ gauss
needed to match the X-ray spectrum, electrons radiating at 10$^{14}$ Hz
have a synchrotron ``half-life'' of $\simeq$ 10$^{4}$ yrs. For a hot spot
radius of $\simeq$ 250 pc, the corresponding streaming velocity of the
relativistic electrons is $\simeq$ 0.1c.
The
problem with model 2 is the fine tuning needed: synchrotron and
      synchrotron self-Compton emission must be present in the Chandra band
      with comparable fluxes.

\subsubsubsection{Inverse Compton and Synchrotron Emission from
Hypothetical New Electron Populations}

In the previous subsection, we showed that the X-ray emission
is so strong that the magnetic field in the hot spot must
be a factor of $\sim$ 14 
below equipartition if 
the X-rays are produced by inverse Compton scattering
from the electrons that generate the {\it observed} radio emission.
Even then,
the observed X-ray spectrum is different to the predicted one
(Fig. 8), and so this model (model 1, Table 4) may be ruled out.
The other process capable of generating the observed X-rays is synchrotron
radiation. However, the X-ray emission of the hot spot is not a
smooth continuation
of the radio -- optical synchrotron spectrum. In model 2, we contrived to
fit the X-ray spectrum by a combination of synchrotron emission from an 
extension of the radio-optical spectrum and synchrotron self-Compton from 
the radio-optical synchrotron-emitting electrons. In view of the
unsatisfactory nature of these models, we now consider inverse Compton and
synchrotron models involving hypothetical new electron populations.

\noindent
{\it i) X-rays as inverse Compton radiation}

For a
successful inverse Compton scattering model, a low energy population
of relativistic electrons with an index of the energy spectrum p
$\simeq$ 2$\alpha_{X}$ + 1 is needed. The required value of p is thus
p$_{{\rm hs}}$ $\simeq$ 3.14$^{+0.22}_{-0.22}$. Fig. 9
shows a model (model 3, Table 4) for the hot spot in which a
synchrotron-emitting component with a larger spectral index than that observed
in the radio has
been added at low radio frequencies. The spectrum of the hot spot has
not been measured below 327 MHz, but the extension of this
hypothetical spectrum down to lower frequencies does not exceed the
measured {\it integrated} flux density of Pictor A (the lowest such
measurements are at $\sim$ 80 MHz, see PRM).
It is
envisaged that the population of electrons which emits the {\it
observed} synchrotron radio emission is in a relatively strong (e.g.
equipartition, Table 3) magnetic field, so that the synchrotron self-
Compton emission from this {\it electron} population is negligible, as
shown above.
However, the observed radio-optical synchrotron {\it radiation}, as
well as the radio synchrotron radiation of the hypothesised component, is 
available for
scattering by the hypothetical low energy electron population. As can
be seen from Fig. 9, the low energy electrons must radiate
predominantly through the inverse Compton channel, requiring a very weak
magnetic field (model 3, Table 4) which is
$\sim$ 100 times weaker than the
equipartition field in the core of the western hot spot.
The calculation shows that the second order Compton scattered component
is $\gtrsim$ 2 orders of magnitude weaker than the first
order Compton scattered component in the Chandra band.
We find
that the spectrum of the low energy electrons has to be steep
(p = 3.3) as shallower spectra (p $<$ 3.3) cause the inverse Compton
  spectrum to increase above $10^{17}$ Hz, contrary to our X-ray
observations.
Taking this model to its extreme, the
hypothetical electron population could be in a region free of magnetic
field, and thus radiate no radio emission at all.  This requirement
for a weak or absent magnetic field associated with this population is
difficult to understand in view of the finding
that the X-ray, optical and radio emitting regions of the western hot spot
are co-spatial (Section 3.2.1 and Fig. 3).
High resolution radio observations at lower frequencies would
  provide
  stronger constraints on this model.

\noindent
{\it ii) X-rays as synchrotron radiation}

Alternatively, the X-ray emission of the hot spot could be synchrotron 
radiation from a separate population of electrons. In such models
(models 4a and b),
the energy index of the X-ray emitting electrons would be
p$_{{\rm hs}}$ $\simeq$ 3.14$^{+0.22}_{-0.22}$, similar to the above
inverse Compton model. In view of the short synchrotron loss times
(a few years for a 5 keV-emitting electron
in the equipartition field), the energy index at injection
would be p$_{{\rm hs, inj}}$ $\simeq$ 2.14$^{+0.22}_{-0.22}$ 
in a steady state, continuous injection model. The synchrotron break frequency
must thus be below the X-ray band but above 10$^{11}$ Hz (otherwise the 
emission would exceed the observed flux at the latter frequency). Models 4b and
4a (Table 4), in which we have adopted the equipartition field,
represent these two extremes for the synchrotron break
frequency, respectively. Thus, in both these models, the extrapolation of the
X-ray spectrum to lower frequencies does not exceed the observed radio,
infrared or optical flux.
Interestingly, the required
value of p$_{{\rm hs, inj}}$
is the same to within the
errors as the canonical index for particle acceleration by a strong
shock (p = 2, e.g. Bell 1978; Blandford \& Ostriker 1978). In this picture,
electrons would
have to be
continuously reaccelerated by such shocks on pc scales.

\subsection{The X-ray Emission of the Jet}

\subsubsection{Radio to X-ray Spectrum}

Unfortunately, there is no information on the radio spectrum of the
jet. We have only PRM's statement that the average excess brightness
of the jet above the background lobe emission is estimated to be about
10 mJy (7\farcs5 beam)$^{-1}$ at $\lambda$ 20 cm. Over the length of
the detected X-ray jet, we then estimate a total 20 cm flux density of
152 mJy, with large uncertainties. The spectral index between 20 cm
and 1 keV is $\alpha_{\rm{rx,jet}}$ = 0.87, which is similar to
$\alpha_{\rm{r}}$ = 0.85, the spectral index for the {\it entire}
radio source for $\nu$ $>$ 400 MHz (PRM). This value of
$\alpha_{\rm{rx,jet}}$ may also be compared with the index within the
Chandra band of $\alpha_{\rm{x,jet}}$ = 0.94$^{+0.43}_{-0.49}$. Given
the errors, it is not out of the question that the jet has a constant
spectral index from GHz frequencies to 10 keV, but the radio
spectrum is needed to check this. Simply joining the radio and X-ray
flux densities with $\alpha_{\rm{rx}}$ = 0.87 implies a total optical
magnitude for the jet of V $\simeq$ 23 mag, and an optical surface
brightness of V $\simeq$ 29 mag (arc sec)$^{-2}$. The fact that the
jet has not been detected in the optical is consistent with these
numbers.

\subsubsection{Non-Thermal Models}

Some estimated parameters of the jet are given in Table 5. It is
apparent that the microwave background radiation dominates the radiant
energy within the jet, unless the radius of the radio
synchrotron-emitting region is much smaller than the observed radius
of the X-ray jet. 
The X-rays from the jet could be either inverse Compton or synchrotron
radiation, which we discuss in turn. Evaluation of these models is limited
by our ignorance of the jet's radio spectrum. For concreteness, we shall
adopt a jet radius of 1$^{\prime\prime}$ (950 pc) for both the radio and
X-ray emission.

\noindent
{\it i) X-rays as inverse Compton radiation}

We consider first inverse Compton models in which Doppler boosting 
is unimportant, and show that such models of the jet's X-ray
emission are implausible. We then consider models including Doppler
boosting.

\noindent
{\it a) Insignificant Doppler Boosting by the Jet}

Adopting $\alpha_{\rm{r,jet}}$ = 0.87, S$_{\rm{jet, 1.4 GHz}}$ = 152 mJy
for the part of the jet detected in X-rays (Section 
4.3.1), and the equipartition field H$_{\rm{jet}}$ =
2.3 $\times$ 10$^{-5}$ gauss,
the predicted inverse Compton scattered X-ray flux falls a factor of 
$\simeq$ 500 below
that observed. In order to match the observed X-ray flux, we need to
reduce the magnetic
field to H$_{\rm{jet}}$ $\simeq$ 7 $\times$ 10$^{-7}$ gauss,
a factor of $\simeq$ 30 below equipartition.
Alternatively, as for the hot spot,
we could invoke a radiation beam from the nucleus to boost the radiation
density in the jet. To obtain the observed X-ray flux from the jet while
retaining the equipartition field requires an (isotropic) nuclear
luminosity of $\sim$ 2 $\times$ 10$^{48}$ erg s$^{-1}$, which is 
implausibly high. We could also suppose, as we did for the hot spot,
that the {\it observed} radio jet contains a stronger
(e.g. equipartition) magnetic field and there is, in addition, a
population of electrons in a weak or absent field. Because we do not 
know the radio spectrum of the jet, and have thus not been able to
demonstrate that the radio and X-ray spectral indices are
different, we cannot prove that the radio- and X-ray-emitting electrons
represent different populations,
as we could for the hot spot.

The following simple fact argues against inverse Compton scattering for the
jet's X-ray emission {\it if Doppler boosting is unimportant}: in the radio 
band, the western lobe dominates
the jet, while in X-rays the converse is true. The ratio of the rates of
energy loss to inverse Compton scattering and synchrotron radiation is

\begin{equation}
{{(\rm dE/dt)_{IC}}\over{(\rm dE/dt)_{synch}}} = {{\epsilon_{rad}}\over{\epsilon
_{mag}}}.
\end{equation}

\noindent
The radiant energy
density in both
the jet and western lobe is dominated by the microwave background
radiation. The equipartition magnetic field in the jet is
$\simeq$ 2.3 $\times$ 10$^{-5}$ gauss (for r$_{\rm j}$ = 1$^{\prime\prime}$,
see Table 5) and that in the lobe is $\sim$ 5 $\times$ 10$^{-6}$ gauss. Thus
one expects that the ratio (dE/dt)$_{\rm IC}$/(dE/dt)$_{\rm synch}$ should be
larger for the lobe than the jet. Given that the lobe's radio
synchrotron radiation overwhelmingly dominates that of the jet, the
lobe's inverse Compton emission should dominate the jet by an even larger
factor, contrary to observation.
However, the jet's inverse Compton emission would be larger if
either a) its magnetic field is well below equipartition, or b) a narrow
beam of radiation is emitted by the nucleus along the jet, providing a
larger $\epsilon_{rad}$ than the microwave background, as discussed
above. Both of these
possibilities are {\it ad hoc} and so we consider the prominence of the jet
compared to the lobe in the Chandra image as an argument against inverse
Compton scattering in the absence of Doppler boosting.

\noindent
{\it b) Significant Doppler Boosting by the Jet}

The above discussion neglects relativistic boosting or diminution by possible
bulk relativistic motion of the jet. There is now a strong case for energy
transport at bulk relativistic velocities to the hot spots in FRII
sources (e.g. Bridle 1996). For emission which is isotropic in the 
rest frame and has a power law spectrum, the observed flux density,
$F_{\nu}(\nu)$,
is
related to the flux density in the rest frame, $F^{\prime}_{\nu}(\nu)$, by

\begin{equation}
F_{\nu}(\nu)\ = \delta^{3 + \alpha}F^{\prime}_{\nu}(\nu)
\end{equation}

\noindent
where $\delta = [\Gamma(1\ - \beta\ \cos\theta)]^{-1}$,
$\Gamma$ is the Lorentz factor of the bulk flow, $\beta$ is the
bulk velocity in units of the speed of light, and $\theta$ is
the angle between the velocity vector and the line of sight 
(e.g. Urry \& Padovani
1995). This equation assumes that the emission comes from a discrete,
moving source. For a smooth, continuous jet, the exponent 3+$\alpha$ becomes
2+$\alpha$ (Begelman, Blandford \& Rees 1984). Equation (6) will describe
relativistic boosting or diminution of the jet's synchrotron radiation, as long
as that radiation is isotropic in the jet's rest frame.

As noted above (Table 5), the radiation density in the jet is dominated
by the microwave background radiation, which is isotropic in the observer's
frame and anisotropic in the rest frame
of the jet. In this case, the principal dependence of the inverse Compton
scattered radiation on $\theta$ is given by (Dermer 1995)

\begin{equation}
F_{\nu}(\nu)\ = \delta^{4 + 2\alpha}F^{\prime}_{\nu}(\nu)
\end{equation}

\noindent
where an additional term which depends slowly on $\cos\theta$ has been
omitted (Begelman \& Sikora 1987; Dermer 1995). Retaining the discrete
source model, the ratio of inverse Compton scattered to synchrotron spectral
flux density is then

\begin{equation}
F_{C,\nu}(\nu)/F_{S,\nu}(\nu) \propto \delta^{1 + \alpha}.
\end{equation}

\noindent
Thus the ratio of inverse Compton to synchrotron flux is increased by
Doppler boosting ($\delta$ $>$ 1).

Our goal is to develop successful models for the jet's X-ray emission for
various assumed angles $\theta$. Selection of a given $\theta$ defines a maximum
value of $\delta$ ($\delta_{\rm max}$ $\simeq$ 1/$\theta$), which is achieved
when
$\Gamma$ $\simeq$ 1/$\theta$. Assuming $\delta_{\rm max}$ (which is the most
optimistic choice, since it minimises the required reduction of H below
equipartition at a given $\theta$), we have computed the value of H required to
reproduce the observed X-ray to radio flux ratio, and compared it with the
equipartition field which would be inferred from the synchrotron emission
by an observer in the rest frame of the jet (estimated by ``deboosting'' the
observed radio flux; H$^{\prime}$ $\sim$ H$^{-0.7}$). The results are given in
Table 6. For our canonical $\theta$ = 23$^{\circ}$ (Section 4.1), the field must
be at least a factor of 6 below equipartition. An equipartition model
demands $\theta$ = 8$^{\circ}$, which is surely too small for a lobe-dominated
FRII radio galaxy like Pictor A (see discussion in Section 4.1). Such an angle
would make Pictor A one of the largest known radio galaxies (length 3 Mpc).
We conclude that inverse Compton scattering of the microwave background is
a viable model for the X-ray emission of the jet of Pictor A, but requires a
magnetic field which is substantially below equipartition.

\noindent
{\it ii) X-rays as synchrotron radiation}

As already noted, the available data are consistent with a single power
law with $\alpha_{\rm{rx,jet}}$ = 0.87 from 1.4 GHz to 10 keV. This is,
however, an unlikely physical situation for a purely synchrotron spectrum
in view of the short energy loss times of the X-ray emitting
electrons to synchrotron and inverse Compton radiation. For electrons
emitting synchrotron radiation at 1.4 GHz and 1 keV in the equipartition
field of 2.3 $\times$ 10$^{-5}$ gauss, the times to lose half their energy
are $\simeq$ 4 $\times$ 10$^{6}$ and 300 yrs, respectively. If the energies of
the radio-emitting electrons are not significantly reduced by synchrotron and
inverse Compton losses, we would expect the radio spectrum to be flatter
than the X-ray spectrum; in the simplest case involving continuous
injection, $\alpha_{\rm{r,jet}}$ = $\alpha_{\rm{x,jet}}$ -- 0.5 $\simeq$ 0.4.
It would thus be valuable to measure $\alpha_{\rm{r,jet}}$.

For a synchrotron model in which the radiation is isotropic in the frame
of bulk jet motion, the observed
ratio of flux
in the approaching side of the jet to the receding side is given by

\begin{equation}
R=\frac{F_{\nu_{\rm app}}(\nu)}{F_{\nu_{\rm rec}}(\nu)}=\left(
\frac{1+\beta\cos\theta}{1-\beta\cos\theta}\right)^{3+\alpha}.
\end{equation}

\noindent
From X-ray observations $\alpha\approx0.9$ and $R\ge10$
at the brightest part of the X-ray jet.
Assuming $\theta$ $>$ 23$^{\circ}$ (section 4.1), then $\beta$ $>$ 0.3.
If the X-rays are the result of inverse Compton scattering of the
microwave background radiation, the exponent 3+$\alpha$ in equation (9)
becomes 4+2$\alpha$ (cf. equation 7) and a somewhat smaller lower limit to
$\beta$ is obtained.

\section{Concluding Remarks}

Our Chandra study of Pictor A has shown that the X-ray emissions from the
jet and the western hot spot are non-thermal. The spectra of both
are well described by an absorbed power law with (flux density) spectral
index $\alpha$ $\simeq$ 1.0. The X-ray spectrum of the hot spot is not a smooth
extension of the radio to optical synchrotron spectrum, which turns down or
cuts off near $\sim$ 10$^{14}$ Hz. Inverse Compton scattering of the
synchrotron 
radio photons by the relativistic electrons responsible for the radio emission
(i.e. a synchrotron self-Compton model) may be ruled out for the hot spot's
X-ray emission, as the predicted spectrum differs from that observed. 

We considered the possible existence of a population of relativistic electrons 
in the hot spot that radiates synchrotron emission at frequencies below that
at which its spectrum has been measured (i.e. $<$ 327 MHz). By choosing an
appropriate index for the energy spectrum of these electrons, we constructed
a successful synchrotron self-Compton model for the X-rays at the price of
reducing the magnetic field to $\sim$ 1\% of equipartition (see Fig. 9 and
Model 3 in Table 4). More generally, relativistic electrons could exist in
regions with weak or absent magnetic fields. Since the properties of
such electron populations are unconstrained by radio observations, one can 
always create inverse Compton models that match the X-ray spectra. It is,
however, then difficult to understand why the X-rays from both the jet and the
hot
spot should correlate so well with synchrotron radio emission, which must
arise in a relatively strong magnetic field. 

The X-ray spectrum of the hot spot may be reproduced in a composite synchrotron
plus synchrotron self-Compton model. In this picture, the spectral index of the
synchrotron radiation is supposed to increase by 0.5 above the break near
10$^{14}$ Hz, as would be expected in a continuous injection model. Addition
of this synchrotron emission to the synchrotron self-Compton emission
expected for a magnetic field a factor of 9 below equipartition reproduces the
observed spectrum (see Fig. 8 and Model 2, Table 4). The model is
contrived, requiring similar fluxes from the two components in the Chandra
band, but cannot be ruled out.

If the jet is non-relativistic, inverse Compton scattering is an 
implausible model for its X-ray
emission, for it requires a magnetic field a factor of 30 below
equipartition. Further, it is hard to understand why the jet is brighter than
the lobe (the opposite of the situation in the radio) in such a
model. If the jet is relativistic, these difficulties are eased, and
we consider inverse Compton scattering by such a jet off the
microwave background a viable mechanism. However, the magnetic
field must still be well below equipartition for plausible angles between
the jet and the line of sight in this lobe-dominated, FRII radio galaxy.

Synchrotron radiation is a plausible X-ray emission process
for both jet and hot spot. Strong, non-relativistic shocks are believed to
accelerate relativistic particles yielding an energy spectrum
n(E) $\propto$ E$^{\rm -p}$ with p = 2 at injection.
Since the half lives of X-ray 
emitting electrons to synchrotron losses are very short ($\sim$ years), the
spectrum steepens and a synchrotron spectral index of $\alpha$ = 1.0 is
expected, in excellent accord with observations (Models 4a, b, Table 4).
The separate population of radio-optical synchrotron emitting electrons
remains unexplained; the radio spectral index -
$\alpha_{r}$ = 0.740 $\pm$ 0.015 - is close to the average for non-thermal
radio sources. Various processes, including acceleration in weak shocks,
synchrotron losses and effects of tangled fields (see summary in Longair 1994,
Ch. 21), have been invoked to account for the difference between the
typical index seen in radio sources and the value $\alpha$ = 0.5 expected
in the canonical model. Hot spots are, of course, associated with two shocks -
an internal, mildly relativistic shock in the jet and a non-relativistic
bow shock in the intergalactic medium. Both may reasonably be expected to
accelerate cosmic rays and the resulting fluxes and energy spectra may differ.

As well as being directly accelerated in shocks, high $\gamma$
relativistic electrons may result from a `proton induced cascade' initiated
by photopion production (e.g. Sikora et al. 1987; 
Biermann \& Strittmatter 1987; Mannheim \& Biermann 1989; Mannheim,
Kr\"ulls \& Biermann 1991). In this process, relativistic proton - photon
collisions create $\pi^{0}$, $\pi^{+}$ and $\pi^{-}$. The last two decay
into relativistic electrons, positrons, neutrinos and antineutrinos
(e.g. Biermann \& Strittmatter 1987). The process thus provides a supply of
synchrotron X- and $\gamma$-ray emitting electrons and positrons
from high energy protons
accelerated by shocks. Mannheim, Kr\"ulls \& Biermann (1991) considered this
process for production of synchrotron X-ray emission in radio galaxy hot
spots. Their calculations suggest that the proton induced cascade produces
X-ray emission about an order of magnitude weaker and with a harder spectrum
than is observed for the western hot spot of Pictor A. As they note, the
predicted luminosity can be increased by increasing the number of relativistic
protons or the number of photons (the latter might occur, for example, if the
hot spot is illuminated by a beam of radiation from the galaxy nucleus).
This process seems promising and further calculations of the expected X-ray 
luminosity and spectrum of a proton induced cascade would be worthwhile.

Our discussion of the western hot spot has assumed it moves outwards
non-relativistically. While this is established for the population of radio hot
spots in general, it is not necessarily true for Pictor A. The fact that
the western hot spot is on the near side and is much
brighter than the eastern hot spots at both radio and X-ray wavelengths
raises the possibility of relativistic outflow. Some of the problems with
inverse Compton models of the hot spot 
might then be eased (as they are for the jet), but the
difference between the radio and X-ray spectral indices remains a
difficulty.

This research was supported by NASA through grant NAG 81027 and
by the Graduate School of the University of Maryland through a research
fellowship to ASW.  We are
extremely grateful to Rick Perley for providing the radio images
published by PRM in numerical form.  We also wish to thank the staff
of the Chandra Science Center, especially Dan Harris and Shanil
Virani, for their help.

\vfil\eject

\clearpage

\begin{deluxetable}{lcccccccc}
\tabletypesize{\footnotesize}
\tablenum{1}
%\footnotesize       % 10 pt
%\scriptsize           8 pt
\tablewidth{0pt}
\rotate
\tablecaption{Spectral Fits to the X-ray Emission of the Western Hot
Spot\tablenotemark{a}
\label{tab:table1}}
\tablecolumns{8}
\tablehead{
\colhead{Model} & \colhead{$N_{\rm H}$} & \colhead{kT} &
  \colhead{Metal Abundance} &
  \colhead{K$^{\rm b}_{1}$} & \colhead{$\Gamma$\tablenotemark{c}} &
  \colhead{K$^{\rm d}_{2}$} & 
  \colhead{$\chi^2$/dof} \\
\colhead{} & \colhead{[$\times$ 10$^{20}$ atoms cm$^{-2}$]} & \colhead{[keV]} &
\colhead{} &
  \colhead{} & \colhead{} &
  \colhead{} & \colhead{} & \colhead{}}
\startdata

Raymond-Smith & $0.0^{+0.2}_{-0.0}$ & $4.8^{+0.6}_{-0.3}$ & solar &
$(3.44^{+0.12}_{-0.12})\times10^{-4}$ & -- & -- & 166 / 97 \\

Raymond-Smith & 4.2 (frozen) & $4.7^{+0.4}_{-0.4}$ & solar &
$(3.9^{+0.1}_{-0.1})\times10^{-4}$ & -- & -- & 232 / 99 \\

Raymond-Smith & $0.0^{+1.3}_{-0.0}$ & $3.5^{+0.4}_{-0.4}$ &
$0.0^{+0.1}_{-0.0}\times$ solar &
$(4.57^{+0.28}_{-0.21})\times10^{-4}$ & -- & -- & 101 / 96 \\
  
Raymond-Smith & 4.2 (frozen) & $2.7^{+0.3}_{-0.3}$ &
$0.00^{+0.04}_{-0.0}\times$ solar & $(5.7^{+0.2}_{-0.3})\times10^{-4}$ &
-- & -- & 122 / 98 \\

Power Law & 7.1$^{+2.3}_{-2.3}$ & -- & -- & -- &
2.07$^{+0.11}_{-0.11}$ & (1.34$^{+0.13}_{-0.11}$) $\times$ 10$^{-4}$ &
95 / 98 \\

\tablenotetext{a}{Errors are 90\% confidence}
\tablenotetext{b}{$K_{1} = 10^{-14} \int n_{e} n_{H} dV/4 \pi D^2$,
  where $D$ is the luminosity distance to the source (cm), $n_{e}$ is
  the electron density (cm$^{-3}$) and $n_{H}$ is the hydrogen density
  (cm$^{-3}$).}
\tablenotetext{c}{$\Gamma$ is the photon index.}
\tablenotetext{d}{$K_{2}$ is the number of photons keV$^{-1}$
  cm$^{-2}$ s$^{-1}$ at 1 keV}
\enddata
\end{deluxetable}

\vfil\eject

\clearpage

\begin{deluxetable}{lcccccccc}
\tabletypesize{\footnotesize}
\tablenum{2}
%\small              % 11 pt
%\footnotesize       % 10 pt
%\scriptsize         %  8 pt
\tablewidth{0pt}
\rotate
\tablecaption{Spectral Fits to the X-ray Emission of the
  Jet\tablenotemark{a}
  \label{tab:table4}}
  \tablecolumns{8}
  \tablehead{
\colhead{Model} & \colhead{$N_{\rm H}$} & \colhead{kT} &
  \colhead{Metal Abundance} &
      \colhead{K$_{1}$} & \colhead{$\Gamma$} &
	    \colhead{K$_{2}$} &
	    \colhead{$\chi^2$/dof} \\
\colhead{} & \colhead{[$\times$ 10$^{20}$ atoms cm$^{-2}$]} & \colhead{[keV]} &
\colhead{} &
\colhead{} & \colhead{} &
\colhead{} & \colhead{} & \colhead{}}
\startdata
Raymond-Smith & 4.2 (frozen) & $3.7^{+1.6}_{-1.1}$ & solar
	      & $(4.4^{+7.6}_{-0.9})\times10^{-5}$ & --    & --    
	    & 27 / 15  \\
Raymond-Smith & 4.2 (frozen) & $2.2^{+1.3}_{-0.6}$ & $(0.0^{+0.3}_{-0.0})\times$
solar
& $(6.7^{+1.3}_{-1.6})\times10^{-5}$ & --    & --  &  19 / 14  \\
Power Law     & 5.8$^{+6.8}_{-5.6}$ & -- & --
 & --      & 1.94$^{+0.43}_{-0.49}$ & (1.45$^{+0.62}_{-0.36}$) 
$\times$ 10$^{-5}$
&  15 / 14 \\
\tablenotetext{a}{Parameters and errors are as defined in the footnotes to
Table 1}
\enddata
\end{deluxetable}
\vfil\eject
\clearpage

\begin{deluxetable}{lc}
\tablenum{3}
%\small              % 11 pt
\footnotesize       % 10 pt
%\scriptsize         %  8 pt
\tablewidth{0pt}
\tablecaption{Parameters of, and within, the Western Hot Spot
\label{tab:table3}}
\tablecolumns{2}
\tablehead{}
\startdata
L(327 MHz - optical)   & $\simeq$ 3 $\times$ 10$^{43}$ erg s$^{-1}$ \\
L(0.1 - 10 keV)        & 4.9 $\times$ 10$^{42}$ erg s$^{-1}$ \\
H$_{eq}$ (bright core) & 4.7 $\times$ 10$^{-4}$ gauss \\
H$_{eq}$ (filament)    & 9.1 $\times$ 10$^{-5}$ gauss \\
$\epsilon_{rad}$ (core, synchrotron rad'n)  & $\sim$ 8 $\times$ 10$^{-12}$ 
erg cm$^{-3}$                                         \\
$\epsilon_{rad}$ (filament, synchrotron rad'n) & $\sim$ 1 $\times$ 10$^{-13}$
erg cm$^{-3}$                                         \\
$\epsilon_{rad}$ (microwave background)        & 4 $\times$ 10$^{-13}$
erg cm$^{-3}$                                         \\
$\epsilon_{rad}$ (galaxy optical light)        & 4 $\times$ 10$^{-16}$
erg cm$^{-3}$                                         \\
\enddata
\end{deluxetable}

\vfil\eject
\clearpage

\begin{deluxetable}{crccccccc}
\tablenum{4}
%\small              % 11 pt
\footnotesize       % 10 pt
%\scriptsize         %  8 pt
\tablewidth{0pt}
\rotate
\tablecaption{Parameters of the Models of the Western Hot Spot
\label{tab:table3}}
\tablecolumns{9}
\tablehead{
\colhead{Model} & \colhead{$n_{e_0}$} & \colhead{$p$} &
\colhead{$\gamma_{\rm min}$} & \colhead{$\gamma_{\rm break}$} &
\colhead{$\gamma_{\rm max}$} & \colhead{Radius} &
\colhead{Magnetic Field} & \colhead{Fraction of}\\

\colhead{} & \colhead{[cm$^{-3}$]} & \colhead{} & \colhead{} &
\colhead{} & \colhead{} & \colhead{[cm]} & \colhead{[gauss]} &
\colhead{equipartition field}}

\startdata

1 & 5.3 & 2.48 &  $8.7\times10^1$ & -- & -- & $7.7\times10^{20}$ &
$3.3\times10^{-5}$ & 0.07 \\

2 & 2.2 & 2.48 & $8.7\times10^1$ & $4.75\times10^5$ & -- &
$7.7\times10^{20}$ & $5.3\times10^{-5}$ & 0.11 \\

3 & 485.0 & 3.30 & $1.0\times10^2$ & -- & $9\times10^3$ &
$7.7\times10^{20}$ & $2.8\times10^{-6}$ & 0.01 \\

4a & $1.0\times10^{-3}$ & 2.14 & -- & $5\times10^{4}$ & -- & 
$7.7\times10^{20}$ & $4.7\times10^{-4}$ & 1 \\

4b & $1.2\times10^{-5}$ & 2.14 & -- & $4\times10^{6}$ & -- &
$7.7\times10^{20}$ & $4.7\times10^{-4}$ & 1 \\
\enddata
\end{deluxetable}

\vfil\eject
\clearpage

\begin{deluxetable}{lc}
\tablenum{5}
%\small              % 11 pt
\footnotesize       % 10 pt
%\scriptsize         %  8 pt
\tablewidth{0pt}
\tablecaption{Estimated Parameters within the Jet\tablenotemark{a}
\label{tab:table5}}
\tablecolumns{2}
\tablehead{
\colhead{}}
\startdata
H$_{eq}$  &  2.3 $\times$ 10$^{-5}$ (r$_{\rm{j}}$/1 arc sec)$^{-4/7}$ gauss \\
$\epsilon_{rad}$ (synchrotron rad'n)$^{b}$  &
$\sim$ 1 $\times$ 10$^{-15}$ (r$_{\rm{j}}$/1 arc sec)$^{-1}$ erg cm$^{-3}$  \\
$\epsilon_{rad}$ (microwave background) &
4 $\times$ 10$^{-13}$ erg cm$^{-3}$                                       \\
$\epsilon_{rad}$ (galaxy optical light) &
$\sim$ 8 $\times$ 10$^{-15}$ (R/1 arc min)$^{-2}$ erg cm$^{-3}$           \\
\tablenotetext{a}{$r_{\rm{j}}$ is the angular radius of the synchrotron
emitting region of the jet, and R is the distance of the region of the
jet from the galaxy.}
\tablenotetext{b}{Assumes an upper cut-off frequency 
of $\nu_{max}$ = 10$^{11}$ Hz for the synchrotron spectrum.
$\epsilon_{rad}$ (synchrotron rad'n) should be multiplied 
by 3.1 if $\nu_{max}$ = 10$^{14}$ Hz and by 11.5
if $\nu_{max}$ = 10$^{18}$ Hz.}

\enddata
\end{deluxetable}

\vfil\eject
\clearpage

\begin{deluxetable}{lcccccccc}
\tablenum{6}
%\small              % 11 pt
\footnotesize       % 10 pt
%\scriptsize         %  8 pt
\tablewidth{0pt}
\tablecaption{Required magnetic field strengths, H, for successful inverse
Compton models of the jet, as a function of the angle, $\theta$,
between the jet and
the line of sight, assuming $\delta$ takes its maximum value,
$\delta_{\rm max}$
\label{tab:table6}}
\tablecolumns{6}
\tablehead{
\colhead{$\theta$} & \colhead{$\delta_{\rm max}$} & \colhead{$\Gamma$} &
\colhead{H} &
\colhead{H$_{eq}$} & \colhead{H/H$_{eq}$} \\
\colhead{} & \colhead{} & \colhead{} & \colhead{(gauss)} &
\colhead{(gauss)} & \colhead{}}
\startdata
Any value$^{a}$& 1  & 1   & 7$\times$10$^{-7}$ & 2.3$\times$10$^{-5}$ & 0.03 \\
23$^{\circ}$   & 2.6& 2.6 & 2$\times$10$^{-6}$ & 1.2$\times$10$^{-5}$  & 0.16 \\
8$^{\circ}$    & 7.2& 7.2 & 6$\times$10$^{-6}$ & 6$\times$10$^{-6}$   & 1 \\
\tablenotetext{a}{Non-relativistic case}
\enddata
\end{deluxetable}
\vfil\eject
\clearpage

\figcaption[fig1.ps]
{Large field image of the X-ray emission from Pictor A, processed as 
described in Section 2.1. Coordinates are for epoch J2000.0.
A smoothing 
of $\sigma$ = 1\farcs0, in addition to that described in Section 2.1, has been 
applied, so
the FWHM resolution is $\simeq$
3\farcs4. Three contours of the 7\farcs5 resolution 20 cm radio image of PRM 
are shown; they
represent 0.025, 0.05 and 0.1 Jy (beam area)$^{-1}$. The following X-ray
features
are noteable: a) the bright X-ray source associated with the nucleus;
b) the 1\farcm9 long X-ray jet extending to the W; c) the western hot spot
4\farcm2 from the nucleus; d) possible faint, diffuse X-ray emission
around the galaxy on the 1$^{\prime}$ scale;
e) faint X-ray emission extending from the
nucleus to the eastern hot spots; f) four compact X-ray sources within
the uppermost radio contour around the eastern hot spot; and g) an X-ray
source near the eastern edge of the frame which, with respect
to the nucleus, is aligned
in a direction precisely opposite to that from the nucleus to the western hot 
spot.
\label{Figure 1}}

\figcaption[fig2.ps]
{Grey scale representation of the full resolution 
Chandra image of the nucleus, jet and western
hot spot of Pictor A. 
\label{Figure 2}}

\figcaption[fig3.ps] {The morphology of the western hotspot of
Pictor~A, at 3.6 cm radio (left panels; PRM),
R band optical (center panels;
R\"oser (1989) and PRM) and X-ray (right panels; this paper)
wavelengths. The upper panels are grey scale representations and the
lower panels are contour plots in which the contours are
logarithmically spaced and separated by a factor of 2. The radio and
optical images have a FWHM resolution $\simeq$ 1\farcs5 (shown as the filled
circles at the bottom right)
and the
X-ray image has a FWHM of 1\farcs2 $\times$ 1\farcs9
(the PSF is shown at the bottom right; no smoothing has been applied to
the X-ray image). The coordinates are for epoch J2000.0
\label{Figure 3}}

\figcaption[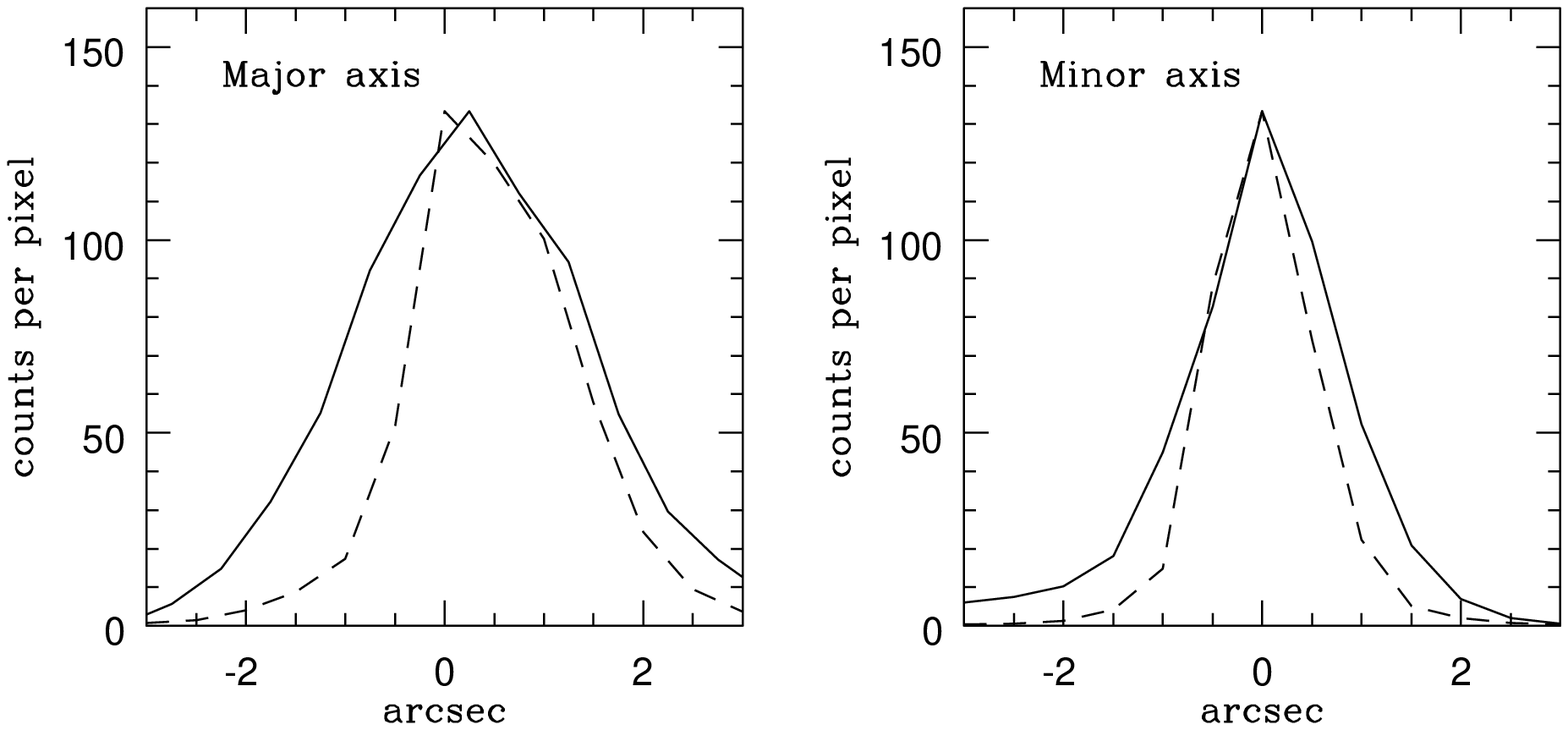] {Profiles of the X-ray emission (solid lines) through the
peak of the western hot spot along its major axis (p.a. $\simeq$ 47$^{\circ}$,
left panel) 
and perpendicular to that direction (right panel). Profiles through
the model psf at
the location of the hot spot in the same directions are shown as dashed lines.
\label{Figure 4}}

\figcaption[fig5.ps] {The X-ray spectrum of the western hot spot of
  Pictor A. The upper panel shows the observed count rate on the
  detector (crosses), and the best model of an absorbed power law
  folded through the instrumental response (solid line). The
  parameters of this model are given in Table 1.  The lower panel
  shows the $\chi$ residuals from this fit.
\label{Figure 5}}

\figcaption[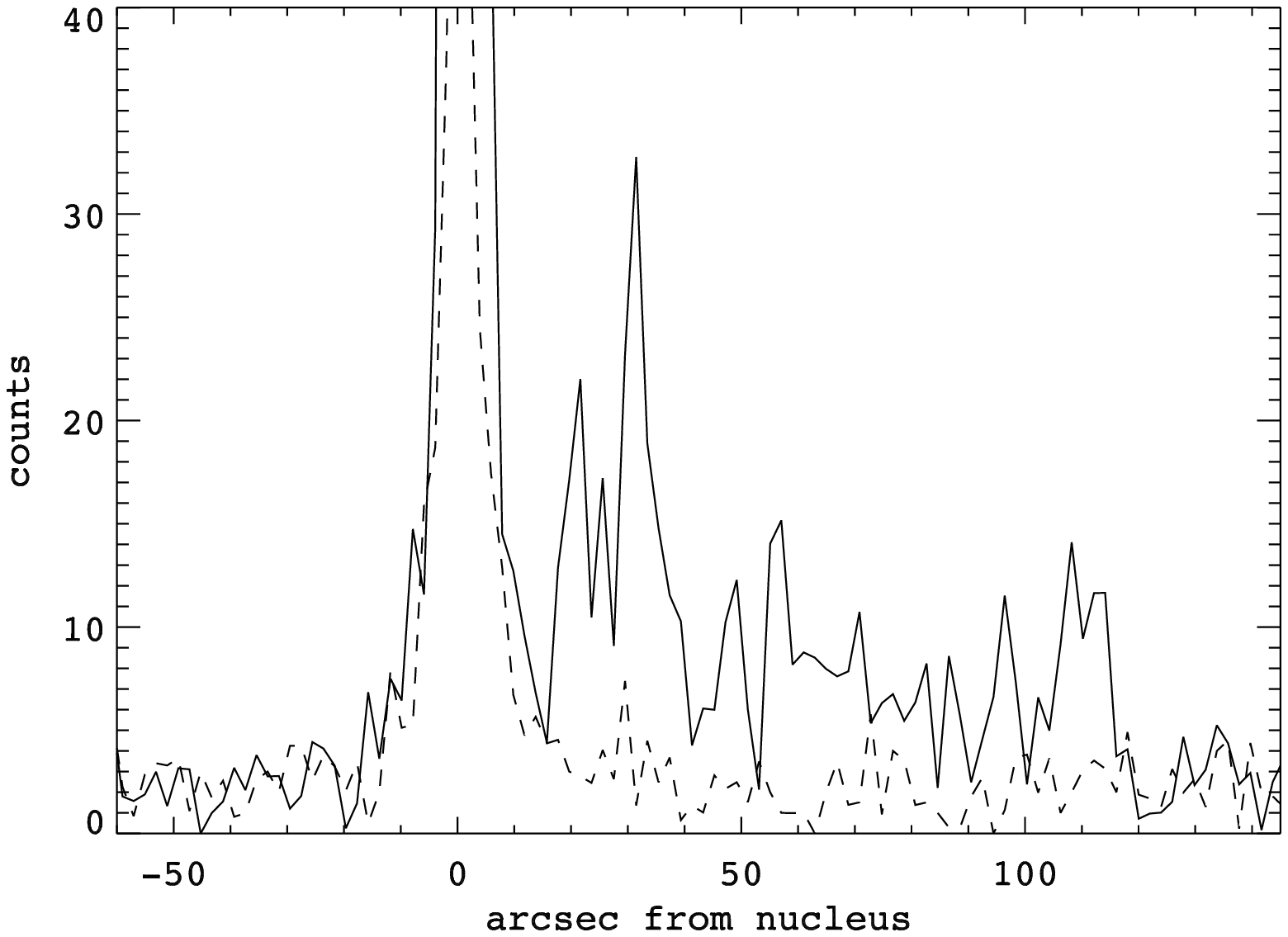] {Profiles of the X-ray emission along the jet
  (solid line) and a background region offset from the jet (dotted
  line). Radial distance from the nucleus increases to the west.
  Counts from the jet were summed over $2\arcsec$ bins in the radial
  direction and $2\farcs5$ bins in the transverse direction. A similar
  region was used for the background, offset from the jet by $2\farcs5$
  in the transverse direction. Excess X-ray emission from the jet is
  clearly seen on the western side of the nucleus out to $1\farcm9$,
  whereas there is no evidence of X-ray emission from a jet to the
  east of the nucleus.
\label{Figure 6}}

\figcaption[fig7.ps] {The X-ray spectrum of the jet of Pictor A.  The
  upper panel shows the observed count rate on the detector (crosses),
  and the best model of an absorbed power law folded through the
  instrumental response (solid line). The parameters of this model are
  given in Table 2.  The lower panel shows the $\chi$ residuals from
  this plot.
\label{Figure 7}}

\figcaption[fig8.ps] {The broad band spectrum, plotted as $\nu S_{\nu}$, 
of the western hotspot of Pictor~A from radio through to
X-ray wavelengths. All the radio points and most of the near
infrared and optical are taken from Meisenheimer, Yates \& R\"oser
(1997). The two optical/near ultraviolet points (represented by
triangles) come from our analysis of archival HST observations of the
hot spot (Section 2.2). The X-ray point with large error bars is
from the Einstein Observatory (R\"oser \& Meisenheimer 1987). The
``bow tie'' represents the spectral range allowed by the
Chandra observations.  The other lines represent models, as
follows. The solid curve called, model 2 in the text, is
the sum of synchrotron emission (passing through the radio data
points and breaking at $\sim10^{14}$\,Hz due to synchrotron losses;
dotted line) and a synchrotron self-Compton component 
(dot-dashed line). The dashed line shows the
synchrotron spectrum expected from a sharp cut-off in the energy
spectrum of the electrons at $\gamma$ $\simeq$ 5 $\times$ 10$^{5}$
in the equipartition magnetic field of 4.7 $\times$ 10$^{-4}$ gauss
(Table 3).
\label{Figure 8}}

\figcaption[fig9.ps]
{The observed spectrum of the western hot spot with the addition of a
hypothetical low frequency ($<$ 327 MHz) radio synchrotron component. Inverse
Compton scattering of the radio-optical photons by the hypothesised low energy
electron population produces an X-ray spectrum (dashed line) 
compatible with the
Chandra-measured spectrum shown as the `bow tie' (model 3; See Section 4.2.2
and Table 4).
\label{Figure 9}}

\clearpage

\end{document}